# Scalable Analysis for Covid-19 and Vaccine Data

Chris Collins, Roxana Cuevas, Edward Hernandez, Reece Hernandez, Breanna Le, Jongwook Woo
Department of Information Systems, California State University Los Angeles
{ccollin2, rcueva12, eherna211, rherna155, ble2, jwoo5}@calstatela.edu

**Abstract:** This paper explains the scalable methods used for extracting and analyzing the Covid-19 vaccine data. Using Big Data such as Hadoop and Hive, we collect and analyze the massive data set of the confirmed, the fatality, and the vaccination data set of Covid-19. The data size is about 3.2 Giga-Byte. We show that it is possible to store and process massive data with Big Data. The paper proceeds tempo-spatial analysis, and visual maps, charts, and pie charts visualize the result of the investigation. We illustrate that the more vaccinated, the fewer the confirmed cases.

## 1. Introduction

COVID-19 has impacted our lives for almost two years. Now that there is a decline in cases, we chose these datasets because it would help better visualize how much the vaccine can positively impact the decrease of Covid. In addition, it will also give us a better analysis and visualization of all general covid cases and the vaccines that are distributed throughout the states.

This topic is essential because it is something that's has been at the forefront of our lives and conversations for the past two years and possibly for the foreseeable future. We can use this data to prove the positive impacts of any findings and analyze any effects we observe to avoid a similar problem in the future. We want our observations of data that is publicly accessible to continue to positively reinforce our peers that vaccines do contribute to the decline of Covid-19 cases. We will look at different types of vaccine datasets, including ones from Pfizer, Moderna, and Janssen, and the data distributed from other states.

The data set collected is about 3.2 GB. Big Data utilizes the data set more efficiently because Big Data is the distributed Parallel Computing system to store and process massive data set. Thus, we adopt Big Data solution using Hadoop and Hive. It is even scalable for data strorage and computation so that the more data set becomes, the more server we can add to the Big Data cluster [13].

## 2. Related Work

This study is a general topic in the world of big data and data analysis. Big Data AI Center (BigDAI) at California State University Los Angeles has completed a study of Covid-19 data [1]. However, its main focus was to highlight disparities in Covid-19 cases amongst people of different races, ethnic backgrounds, and age groups. BigDAI's work solely focused on the confirmed and fatality cases in the world. Although this is an exciting topic that assists with local government decision-making, we expand our analysis and see more pictures and how Covid-19 cases and the rollout of vaccines impact the entire US population. We, too, were able to highlight the disparities in positive covid-19 cases, but more importantly, we could do more with the vaccination data.

Global Banking News also reported on a study by The World Health Organization (WHO) that says fewer COVID-19 cases with more vaccination. Our study also shows that as more vaccines were rolled out, there was a decline in reported cases of COID-19. With our Big Data analysis, we can see an evident decline in the number of cases and deaths compared to in 2020. We see a correlation between an increase in vaccines and a decrease in the confirmed and fatality cases. WHO also states it. This paper enforces our study to show the same results.

An article by The Washington Post covers the topic on the effects of COVID-19 on men and women. This article is very informative about possible reasons why the coronavirus could affect men more severely than women. The article explains that women have a stronger immune system compared to men's immune system. Another reason highlighted in the article is the different social norms for men and women, alluding to the fact that men tend to take it less seriously than women. In the paper, we show the number of cases for men and women in each state.

### 3. Hardware and Data Specifications

The collected dataset consists of the different types of vaccines and their distribution state. It also includes data about the first and second doses. We have taken data from the CDC website because it was the most accurate and updated dataset, and it is open to the public. The dataset size is 3.21GB, and we have taken data from the dates (January 1, 2020 - April 30, 2021). Table 1 shows the data set files and their sizes.

*Table 1 Data Specification*

| Data | Size (Total 3.21 GB) |
|---|---|
| COVID-19_Case_Surveillance_Public_Use_Data_with_Geography | 3,365,639 KB |
| COVID-19_Vaccine_Distribution_Allocations_by_Jurisdiction_-Janssen | 12 KB |
| COVID-19_Vaccine_Distribution_Allocations_by_Jurisdiction_-Moderna | 40 KB |
| COVID-19_Vaccine_Distribution_Allocations_by_Jurisdiction_-Pfizer | 42 KB |
| State_Vaccine_Totals | 5 KB |
| us-daily-covid-vaccine-doses-administered | 220 KB |

The table below shows the specifications for the Big Data cluster given by the Big Data center at SCU (Seoul Cyber University) in Korea. The Big Data cluster provides Hadoop and its ecosystems such as Hive, Pig, Spark and Sqoop.

*Table 2 Hardware Specifications at SCU*

| Number of Nodes | 3 |
|---|---|
| Memory | 192GB |
| Storage | 24TB/18TB RAID |
| OCPU's | 12 |
| CPU speed | 3GHz |

### 4. Workflow

Figure 1 shows the workflow chart with the steps to obtain and analyze the data. First, we download the raw data of COVID-19 from the CDC website and save them to HDFS (Hadoop Distributed File Systems) at the Big Data cluster. We create several tables with columns to afford the datasets.

We transform data to fit to the table. The data transformation includes adjusting the States columns to a consistent spelled out State name rather than an abbreviated format. A similar data transformation was done for other datasets that contain a different date format than what we needed to avoid any lack of standardization among the fields collected. We also join some tables for the data analysis. We created a directory and individual files where each dataset can be stored. This allowed us to stay organized and accurately query each table, as having multiple datasets within one folder would query only the first available file, resulting in erroneous data.

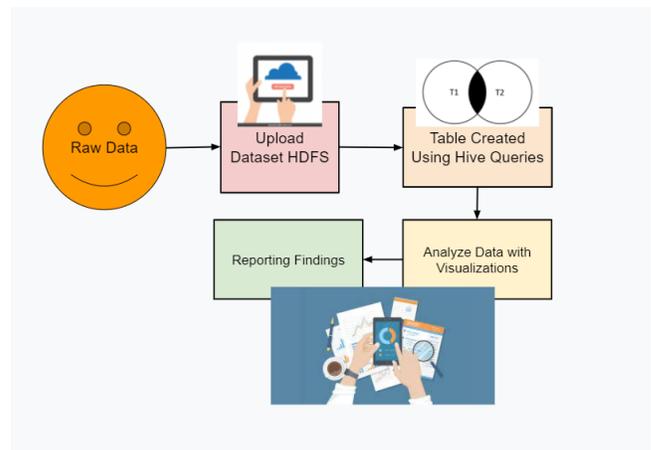

*Figure 1 – Workflow Chart*

After we created tables, we developed Hive codes to pull the information such as the total vaccines administered, the number of the fatality cases, the total of vaccines distributes, and breakdowns of this data by the states. Lastly, we visualize the result data with query outputs on bar graphs, pie charts, and geo maps.

## 5. Data Cleaning

We used Excel for some data cleaning and uploaded the finalized data files into the HDFS. The data clean is to make the standard format for the vaccines from Moderna, Pfizer, and Janssen. We can analyze the total vaccination by the state with the confirmed cases. Using Excel, we changed the state columns to the format of full name, which helps accurately join across all tables. For example, we change CA to California, NY to New York, TX to Texas, etc. Additionally, the raw datasets contain date formats that needed to be adjusted to match across the Hive tables. We transform the dates formats to MM/DD/YYYY.

## 6. Analysis and Visualization

Here are analyses and visualization that we put together from Tableau, Power BI, and Excel. We used different charts to show the relationship between the cases and the vaccines. Our visualizations consist of the confirmed cases in California and by the state in the US, Pfizer Vaccine distribution in the US, the Death Count, and the Vaccination impact on the confirmed cases.

reporting in 2020 and there was a few cases in California. The confirmed cases are higher than in most months of 2021 except April. The data for April is significant because in 2020, California started lockdown. And, in April 2021, vaccines became available to all persons over 18. The chart in that month shows us that vaccines have positively affected on decreasing the confirmed cases of COVID-19.

Figure 3 displays a map of the United States showing the confirmed cases as red dots. The larger the surface area of the circles, the larger the population of the confirmed cases in that state. The states with the highest number of cases in order are California, New York, and Florida.

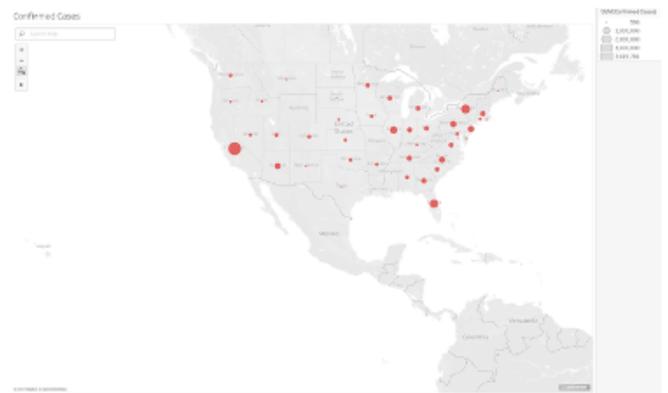

*Figure 3 – Confirmed Cases by State in the United States*

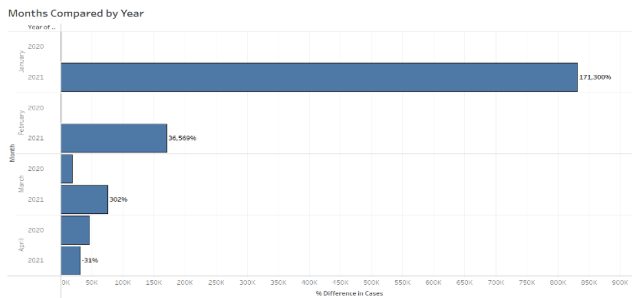

*Figure 2 – Comparison of Covid-19 Cases in California for 2020 and 2021*

### 6.1 Vaccination and the Confirmed Cases Analysis in Tableau

Figure 2 compares the confirmed cases in California from 2020 to the same month in 2021. There is a significant difference in January from year to year. In the beginning, it had poor

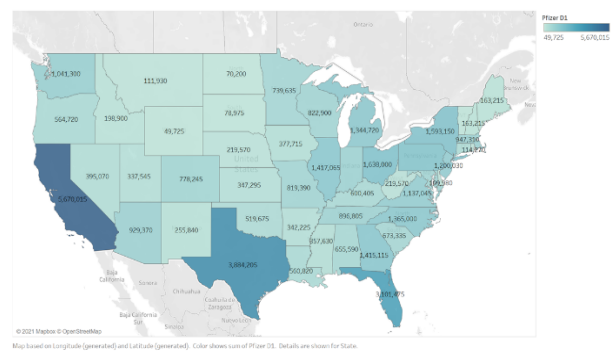

*Figure 4 – Pfizer Vaccinations – the most distributed vaccine for the first dose.*

Figure 4 represents the distribution of the first dose Pfizer vaccine in the United States. It displays both count and state of distribution. The distribution of Moderna and Janssen shows no deviations from the trend. The darker the color means the more distributions in the state. The top three states with the most distributions are California with 5,670,015 distributions, in second place, 3,884,205 distributions in Texas, and lastly, Florida with 3,101,475 distributions.

Figure 5 describes the Total fatality cases by the month. It illustrates the rise of Covid cases peaked in April of 2020. The cases started decreasing during May 2020 and Oct 2020. It should be because of the states' staying at home order. The confirmed cases rose again, approaching the holidays towards the end of 2020. After we see the drastic drop as the month of the vaccines are distributed starting in late Dec 2020- early Jan 2021.

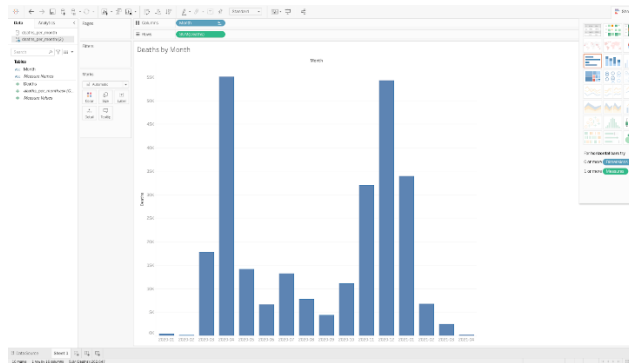

*Figure 5 – Total Fatality cases*

**6.2 Vaccination and Gender analysis in Power BI**

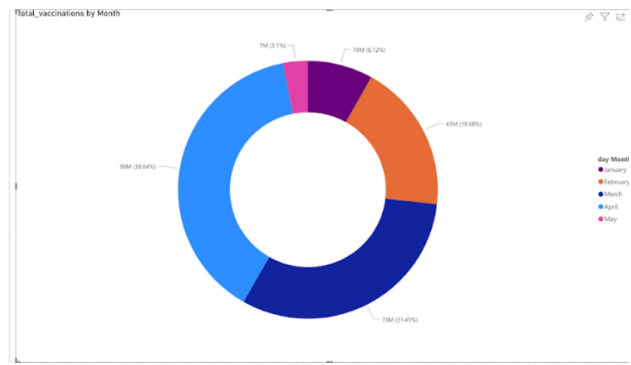

*Figure 6 – Total number of Covid-19 vaccines administered by month*

The total number of vaccines administered in months is visualized below using Power BI. We can observe the most extensive vaccine distribution during March and April of 2021, in which over 232M vaccines were administered.

Most states show a slightly higher number of positive COVID cases for females than males except Texas. In regards to Texas, males tested positive for COVID-19 more frequently than females.

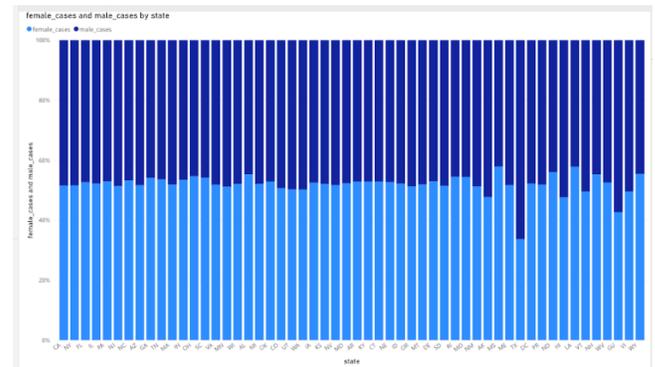

*Figure 7 – Female Vs Male Cases by State*

**6.3 Impact of Vaccination in Excel**

Figure 8 shows the correlation between cases and total vaccinations in a line graph. As total vaccinations increase, the number of confirmed cases decreases. Since vaccinations just started in December 2021, cases are declining slowly but surely till April 2021 as more people take on getting vaccinated.

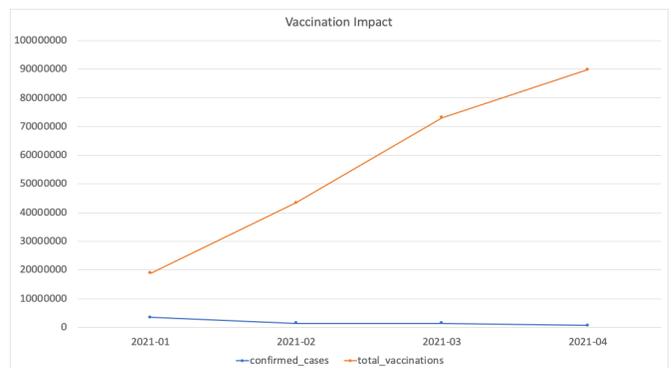

*Figure 8 – Vaccination Impact on Confirmed COVID Cases*

**5. Conclusion**

Based on the experimental result as of April 2021, we can conclude the following:

1. As covid vaccinations increase, covid cases decrease.
2. The confirmed cases were highest in January 2020 to the beginning of 2021 compared to now in California.
3. The states with the highest confirmed cases in the United States were California, New York, and Florida.
4. The most distributed vaccine in the United States is Pfizer.
5. The total fatality cases rose and declined throughout the months due to the state orders, holidays, and vaccines administrations.
6. On average, 51%-54% of females in most states contracted Covid-19 more often than males. The only exception to this was Texas, where only 33% of positive cases were females.

We analyze 3.2 GB COVID-19 data sets from the beginning to April 2021 using the scalable Big Data. Understandably many organizations today rely on this type of information for future operations. Yet, Big Data should be more efficient for the large data sets on a country and worldwide scale.

Processing Techniques and Applications (PDPTA), Las Vegas. 2011.